\documentstyle[12pt,aasms4]{article}

\slugcomment{Subtmitted to ApJL, 2/19/96}

\lefthead{Teegarden et al.}
\righthead{TGRS Observation of Galactic Center}

\begin{document}

\title{TGRS Observation of the Galactic Center\\
    Annihilation Line}

\author{B. J. Teegarden\altaffilmark{1}, T. L. Cline, N. Gehrels,
   D. Palmer, R. Ramaty, H. Seifert,}
\affil{NASA/Goddard Space Flight Center}

\author{K. H. Hurley}
\affil{University of California, Berkeley}

\author{D. A. Landis, N. W. Madden, D. Malone, R. Pehl }
\affil{Lawrence Berkeley Laboratories}

\and

\author{A. Owens}
\affil{University of Leicester}

\altaffiltext{1}{On leave at Centre d'Etudes Spatiale des Rayonnements, 
     Toulouse}

\begin{abstract}

The TGRS (Transient Gamma-Ray Spectrometer) experiment is a high-resolution 
germanium detector launched on the WIND satellite on Nov.  1, 1994.  
Although primarily intended to study gamma-ray bursts and solar flares, 
TGRS also has the capability of studying slower transients (e.g.  x-ray 
novae) and certain steady sources.  We present here results on the narrow 
511 keV annihilation line from the general direction of the Galactic Center 
accumulated over the period Jan.  1995 through Oct.  1995.  These results 
were obtained from the TGRS occultation mode, in which a lead absorber 
occults the Galactic Center region for 1/4 of each spacecraft rotation, 
thus chopping the 511 keV signal.  The occulted region is a band in the sky of 
width 16\arcdeg \ that passes through the Galactic Center.  We detect the 
narrow annihilation line from the galactic center with flux = 
$(1.64\pm0.09)\times10^{-3} \
\mbox{ph} \ \mbox{cm}^{-2} \
\mbox{s}^{-1}$.  The data are consistent with a single point source at 
the galactic center, but a distributed source of extent up to 
$\sim$30\arcdeg \ cannot be ruled out.  No evidence for temporal 
variability on time scales longer than 1 month was found.

\end{abstract}
 
\section{Introduction}

Electron-positron annihilation radiation is a valuable diagnostic for 
studies of a variety of different types of astrophysical sources.  
Positrons can be produced by hot pair plasmas in the vicinity of compact 
objects, by nucleosynthetic processes (both hydrostatic and explosive) and 
by cosmic-ray interactions in the interstellar medium.  A narrow 511 keV 
electron-positron annihilaton line ($\sim$ 3 keV FWHM) has been observed 
from the general direction of the Galactic Center (see \cite{tee94} and 
\cite{ram95} for recent reviews).  Data from a series of balloon 
observations in the late seventies and eighties suggested that the line was 
variable.  These same instruments had a variety of different 
fields-of-view, and it was noted (see \cite{tee94}) that the detected flux 
varied systematically with the size of the field-of-view.  This implied 
that there was also a diffuse steady component to the narrow 511 keV 
emission.  Subsequent satellite observations with SMM (\cite{sha90}) and 
CGRO/OSSE (\cite{pur93}, 1994) found no evidence for variability.  CGRO/OSSE has 
now mapped the Galactic Center region with its 3\arcdeg \ $\times$ 
11\arcdeg \ collimator (\cite{pur93},1994).  The OSSE data can be 
well-represented by a two-component distribution consisting of a central 
bulge component of width $\sim8\arcdeg$ \ combined with a disk-like 
component.  $\sim75\%$ of the total flux is contained in the bulge 
component and $\sim 25\%$ in the disk component.

The GRANAT/SIGMA experiment reported episodes of transient broad 
line emission, lasting typically a fraction of a day, from the source 
1E1740.7-2942, which lies within 1\arcdeg \ of the Galactic Center 
(\cite{bou91}; \cite{sun91}; \cite{cor91}).  The line energy was $\sim480$ 
keV, and it was broadened to $\sim130$ keV.  The features were interpreted as 
electron-positron annihilation radiation redshifted by the strong gravitational field 
of a black hole (\cite{bou91}; \cite{sun91}).  It has been suggested 
(\cite{ram92}) that sources such as these can inject positrons into the ISM 
which can travel substantial distances before annihilating.  An ensemble of 
such sources could be responsible for the diffuse narrow 511 keV emission 
from the direction of the Galactic Center.

\section{Instrumentation}

TGRS (Transient Gamma-Ray Spectrometer) is an unshielded high-resolution 
35-cm$^2$ germanium detector passively cooled to a temperature of 85K (for 
more detail see \cite{owe95}).  Its energy range is $\sim 20 - 8000$ keV, 
and the energy resolution after launch was $\sim2.7$ keV FWHM at 500 keV.  
TGRS was launched on the WIND spacecraft on 1 Nov.  1994.  The primary 
purpose of the WIND mission is to sample conditions in interplanetary space 
in the vicinity of the earth's magnetosphere.  Although in earth orbit, 
WIND spends the great majority of the time in interplanetary space.  As a 
consequence the environment is generally quite stable (background levels, 
temperatures, etc.) which lends itself quite well to the long data 
accumulations necessary for the analysis described in this paper.  A 
schematic view of the detector/occulter assembly is shown in Figure.  1.  
The main axis of the detector is coincident with the spin axis of the 
spacecraft, which is normal to the plane of the ecliptic.  The +Z axis of 
the spacecraft (and the TGRS main field-of-view) is oriented towards the 
south ecliptic pole.  This orientation was chosen so that the Large 
Magellanic Cloud (presumed location of the 5 Mar 1978 gamma-ray burst) is 
in the TGRS field-of-view.  The occulter is a 1-cm thick lead absorber 
located just outside of the radiative cooler in the plane of the detector, 
which subtends an angle of 90\arcdeg \ with respect to the detector.  
Additional structure and shielding (not shown in Figure 1) limit the 
low-energy threshold in the occultation mode to $\sim40$ keV.  The on-board 
processor accumulates data in 4 commandable energy windows (64 channels 
ea.) synchronized with the spin of the spacecraft.  One of these is 
centered on the 511 keV annihilation line.  Each spacecraft rotation 
(period = 3 s.) is divided into 128 sectors.  Due to spacecraft telemetry 
limitations one complete readout of all spin-synchronized data takes 
$\sim2$ hr, which defines the minimum time resolution in this mode.

Figure.  2 shows the region of the sky in galactic coordinates that is 
occulted (FWHM = 16\arcdeg) by TGRS.  The occulter has been offset by an 
angle of 5.5\arcdeg \ with respect to the spin plane to make the center of 
the occultation band coincide with the Galactic Center.  The Crab also 
lies within the occultation region.  Also shown in Figure 2 is a contour 
plot of the OSSE 511 keV flux distribution (\cite{pur94}).  As can be seen, 
the TGRS occulter modulates most of the flux in the central bulge component 
of the OSSE distribution.

\section{Observations}

Preliminary data are presented here for the period Jan.-Oct. 1995.  The 
excellent stability of the instrument allows accumulations of long periods 
of data with no requirement for gain corrections.  The 511 keV background 
line has been used for gain calibration.  Figure 3a shows the counting 
rate in a window 10 keV wide centered on the 511 keV line vs.  the spin 
phase of the spacecraft, which is divided into 128 bins.  The on-board 
accumulation is synchronized by a sun-sensor, and the data have been 
corrected for the sidereal variation of the phase.  The solid line is the 
best fit to the data ($\chi^{2} = 155$, 125 degrees of freedom, 
$P(\chi^{2},\nu)=0.036$) for a single point source model in which the 
amplitude and position of the source are allowed to vary.  The best-fit 
value for the source flux is $(1.64\pm0.09)\times10^{-3} \ \mbox{ph} \ 
\mbox{cm}^{-2} \
\mbox{s}^{-1}$ and for the sector no. of the source is $95.3\pm0.4$.  The sector
number corresponding to the Galactic Center is 94.9.  To check for 
systematic effects that could mimic the signal from the Galactic Center a 
control region (516 - 539 keV) adjacent to the 511 keV window was also 
examined.  This is shown in Figure 3b.  There is no evidence for any 
significant signal from the direction of the Galactic Center.  A small 
residual in the sector range ~ 15 - 40 is present due to hard x-ray 
emission from the Crab.

The data have been tested for the possible presence of a broadened source.  
Two broadened models (rectangular and guassian profiles in ecliptic 
longitude) were fit to the data.  While the best fit values for the widths 
were non-zero ($21\arcdeg$ in the rectangular case and $16\arcdeg$ in the 
gaussian case) the reduction in $\chi^{2}$ was not significant 
($\Delta\chi^{2}
\approx 2$, 124 degrees of freedom, for both models).  The 90\% confidence upper
limit on the width of the rectangular model is 34\arcdeg and on the FWHM 
of the gaussian model is 30\arcdeg.  {\it The TGRS data are therefore 
consistent with a point source at the Galactic Center, however, an 
extended source distribution with width up to $\sim30\arcdeg$ cannot be 
ruled out.\/}.

Figure 4 shows the channel-by-channel spectrum in the vicinity of 511 keV 
derived from the TGRS occultation data.  The spectrum was calculated for an 
assumed point source at the Galactic Center.  The data over the 300 day 
period were accumulated as a function of energy and spin phase.  The 
spectra were derived as follows.  Let $C_{i}(E_{j})=$ no. of observed 
counts in the i$^{th}$ spin sector and the j$^{th}$ energy channel, 
$M_{i}=$ transmission of occulter for the i$^{th}$ spin sector, $S(E_{j})=$ 
source counts in the j$^{th}$ energy channel and $B(E_{j})=$ background 
counts in the j$^{th}$ energy channel.  Over the limited band of energy in 
question (506 - 516 keV) $M_{i}$ is taken to be independent of energy.  The 
counts in each sector and energy bin are then given by:
\begin{equation}
C_{i}(E_{j}) = M_{i}S(E_{j}) + B(E_{j}) 	
	\label{}
\end{equation}
The source spectrum is derived by linear regression on $C(E_j)$ of M, with
a constant term.  The coefficient of M gives the source flux and the
constant B gives the background at $E_j$.
This is equivalent to the previously discussed model fit with the
source location held fixed (at the best fit value) and the energy range restricted
to a single energy channel. 

The spectrum in Figure 4 displays a narrow line at 511 keV.  A gaussian 
fit to the line yields a flux of $(1.64\pm0.08)\times10^{-3} \ \mbox{ph} \ 
\mbox{cm}^{-2} \ \mbox{s}^{-1}$ in excellent agreement with that derived 
from the data of Figure 3.  There is a suggestion of line 
broadening. However, no gain shift corrections have 
been applied to the data, and there may, as a reult, be some broadening 
beyond the normal instrumental resolution.  Further careful analysis 
will be necessary to remove this effect (and possibly others) to obtain a 
reliable absolute value for the line width.

Figure  5 is a time history of the intensity of the Galactic Center 511 keV line plotted 
on a monthly basis.  The data points were derived by determining the 
best-fit values for the flux using the TGRS spin sector data such as shown 
in Figure  3.  A single point source was assumed, and both the flux and 
position were allowed to vary.  In all cases the best-fit position was 
consistent with Galactic Center.  The data are consistent with a 
constant flux with a $\chi^{2}$ probability of $\sim17\%$. 

\section{Discussion}

Through its occultation mode TGRS has detected a significant 
($\sim20\sigma$) flux of 511 keV emission from the direction of the 
Galactic Center.  The region observed corresponds to a band in the sky of 
FWHM 16\arcdeg \ offset by -5.5\arcdeg \ from the ecliptic plane, which 
passes through the Galactic Center (see Figure 2).  Purcell et al. (1994) 
have fit the OSSE 511 keV data with a diffuse distribution consisting of a 
central bulge component and a disk component (see Figure 2).  The 
respective fluxes in these two components are $1.7\times10^{-3} \ 
\mbox{ph} \ \mbox{cm}^{-2} \
\mbox{s}^{-1}$ and $0.5\times10^{-3} \ \mbox{ph} \ \mbox{cm}^{-2} \
\mbox{s}^{-1}$.  If one convolves the OSSE model 
distribution with the TGRS occulter modulation function (for the occulter 
centered at the Galactic Center) one obtains an occulted flux of $1.5\times10^{-3} \ \mbox{ph} \ \mbox{cm}^{-2} \
\mbox{s}^{-1}$.  This is to be compared with the measured TGRS occulted 
flux of $(1.64\pm0.09)\times10^{-3} \ \mbox{ph} \ \mbox{cm}^{-2} \
\mbox{s}^{-1}$. The TGRS and OSSE absolute flux measurements are 
therefore mutually consistent at approximately the $1\sigma$ level.

The TGRS data can be well fit by a point source at an ecliptic longitude of 
$268.0\pm1.2\arcdeg$ (J2000) which is $1.0\sigma$ from the Galactic Center 
( ecliptic longitude = 266.8\arcdeg).  The ecliptic longitude of the source 
1E1740.7-2942 is 266.4\arcdeg \ or within $1.3\sigma$ of the best-fit TGRS 
position.  The TGRS 511 keV source position is therefore consistent with 
both the Galactic Center and the 1E1740.7-2942 source.  Because of the 
proximity of the TGRS occulter to the germanium detector, and the finite 
size of the detector itself, the effective TGRS point spread function is 
rather large (see Figure 3).  It is therefore not possible to make a 
definitive statement on the spatial extent of the source.  The data were 
fit with two different broadened source distributions, rectangular and 
gaussian.  While the best-fit values for the width were non-zero (21\arcdeg 
\ for rectangular and 16\arcdeg \ FWHM for the gaussian distribution), the 
change in $\chi^{2}$ was not significant ($\Delta\chi^{2}\simeq2$ in both 
cases).  The 90\% confidence upper limits on the widths were 34\arcdeg \ 
for the rectangular model and 30\arcdeg \ FWHM for the gaussian model.

The historical evidence for variability in the Galactic Center 511 keV flux 
comes primarily from a series of balloon flights from the late 1970's to 
the early 1990's.  Two flights in 1981 (\cite{lev82}; \cite{pac82}) and one 
in 1985 (\cite{lev86}) found no detectable narrow 511 keV line flux.  These 
instruments had fields-of-view in the 15-20\arcdeg \ range and sufficient 
sensitivity to detect emission at the levels implied by the OSSE data.  
Teegarden (1994) showed that these data were inconsistent with a constant 
flux at the 1\% level.  However, the error estimates generally did not 
include systematic effects, which could reduce the significance of the time 
variation.  Initial results from HEAO-3 suggested a variation in the 
Galactic Center 511 keV flux between the fall of 1979 and the spring of 
1980 (\cite{rie81}).  However, a subsequent re-analysis by Mahoney et al.  
(1993) reduced the variation to an insignificant value.  None of the other 
satellite observations (SMM, OSSE, TGRS) have found any evidence for 
temporal variability.  The available body of evidence weighs most 
heavily in favor of a constant diffuse source for the narrow Galactic 
Center annihilation radiation, although a variable point source at a level 
$< 2\times10^{-4} \ \mbox{ph} \ \mbox{cm}^{-2}$ cannot be ruled out 
(\cite{ram94}; \cite{pur94}).
\acknowledgments 

The authors would particularly like to recognize the efforts of the 
ISTP/WIND project staff (J.  Hrastar, project manager, W.  Anselm, 
instrument coordinator).  Particular credit is to be given to D.  A.  
Sheppard, D.  Stilwell, A.  Post, S.  Matthias, C.  Cork and P.  Luke for 
the development of the flight hardware.  Special recognition is due to S.  
Bansal and T.  Sheets for the development of the TGRS data processing 
software.  The authors wish to thank Wm.  Purcell for many helpful 
comments.

\clearpage

\figcaption[fig1.ps]{Schematic view of the TGRS detector/occulter.  The 
spacecraft spin axis is coincident with the detector axis and normal to the 
ecliptic plane.  The occulter subtends an angle of $90\arcdeg \times 
16\arcdeg$ (FWHM) with respect to the detector.}

\figcaption[fig2.ps]{Band of the sky modulated by the TGRS occulter.
Also shown is a contour plot of the 511 keV annihilation radiation as 
measured by OSSE.  Contours are logarithmically spaced with 
a ratio between contours of 3.5.  The location of the Crab, which lies 
within the occulation band, is also shown.}

\figcaption[fig3.ps]{Counting rate as a function of spacecraft spin 
sector (128 sectors/spin).  Data are accumulated between 1 Jan.  1995 and 
27 Oct.  1995 (300 days).  (a) Energy range 506 - 516 keV.  Solid line is 
best-fit single point source model.  The location is consistent with 
Galactic Center.  (b) Energy range 516-539 keV.  No evidence for a 
significant source at the Galactic Center is present.}

\figcaption[fig4.ps]{TGRS spectrum in vicinity of 511 keV line.  Data 
accumulated from 1 Jan. 1995 to 27 Oct. 1995.  Energy of line is 
consistent with 511 keV. }  

\figcaption[fig5.ps]{Time history of 511 keV line intensity during 1995.}

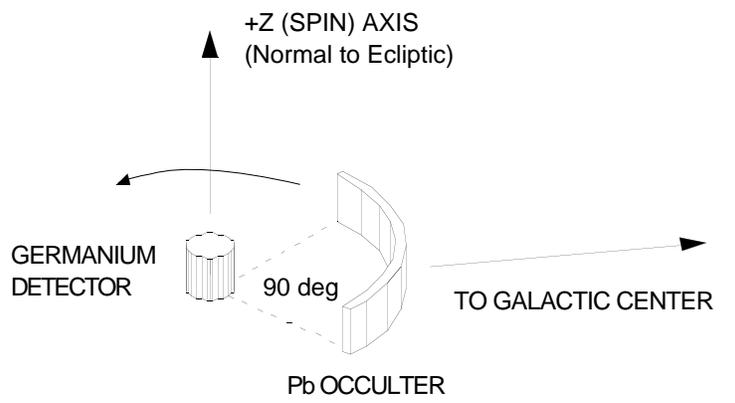
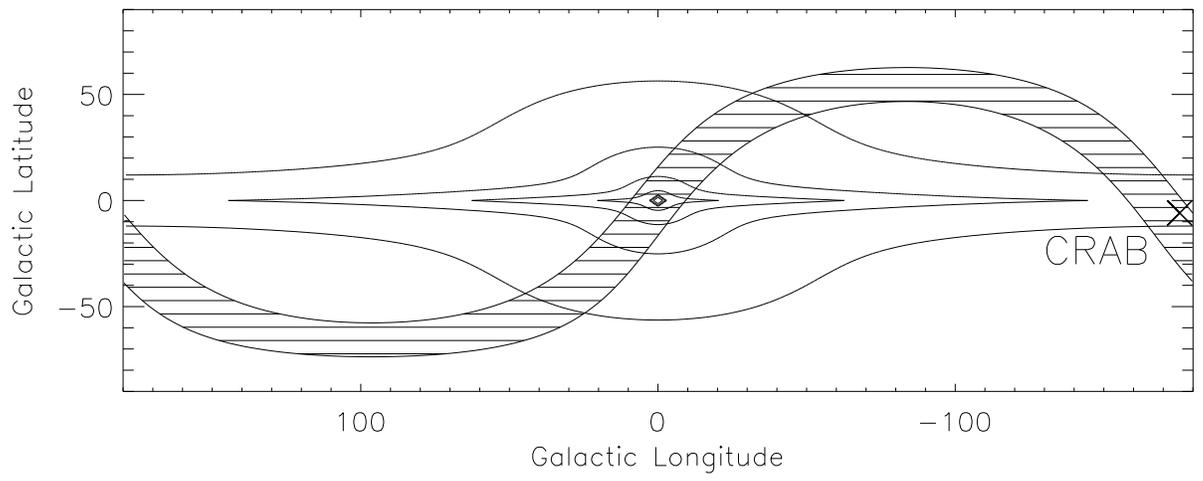
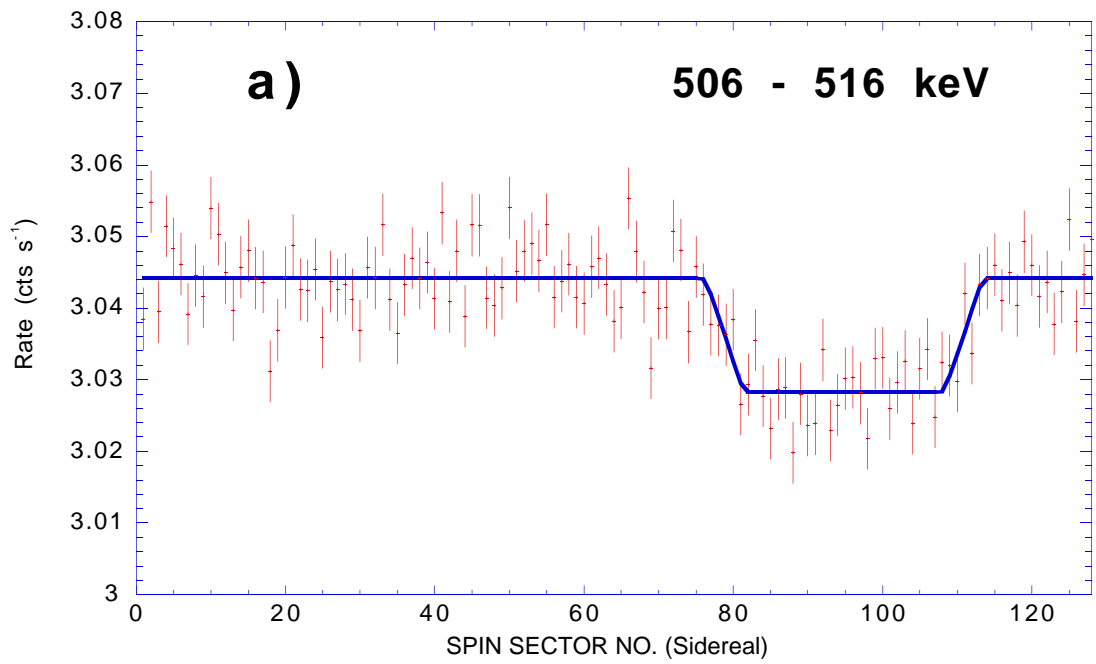
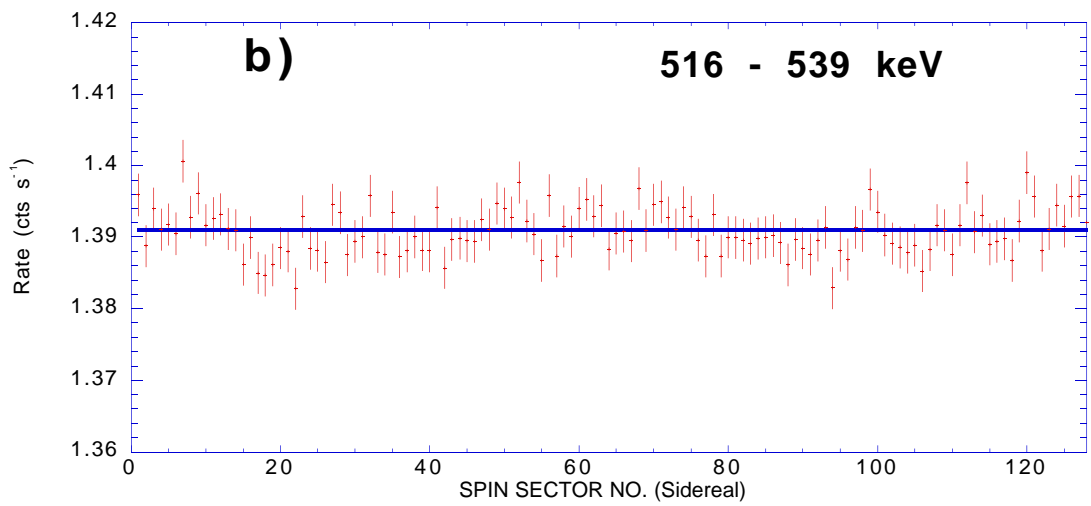
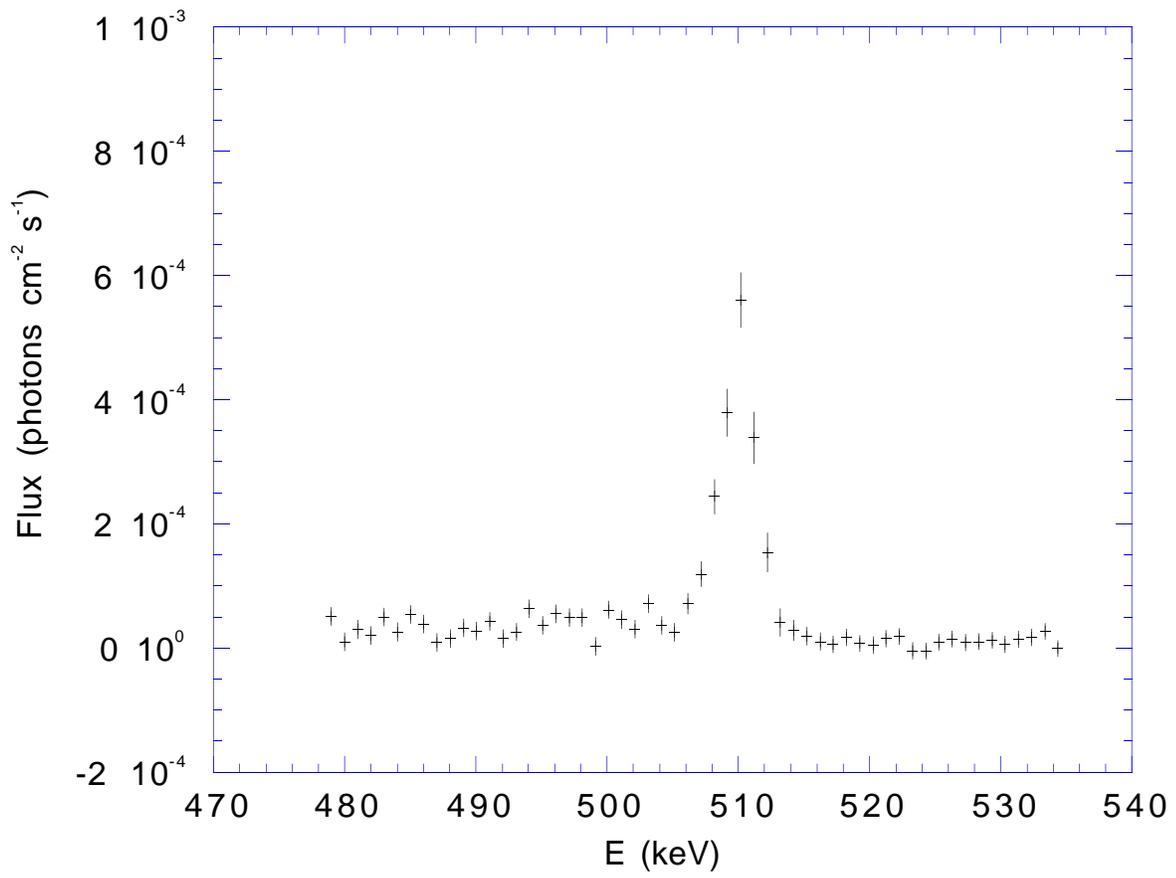
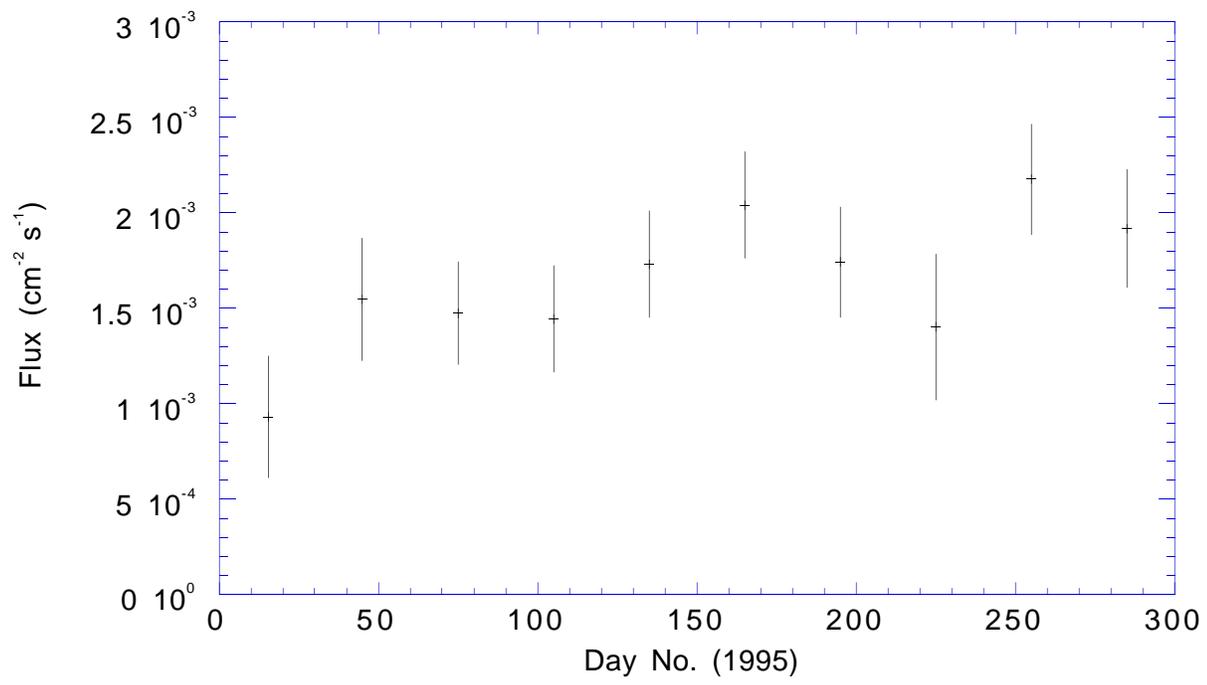
\end{document}